\documentclass[11pt,twoside]{article}
\usepackage{asp2010}

\usepackage{amsmath}

\newcommand{\Teff}{\ensuremath{T_\mathrm{eff}}}

\newcommand{\finfty}{\ensuremath{f_\infty}}
\newcommand{\vel}{\ensuremath{\varv}}
\newcommand{\velcl}{\ensuremath{\vel_\mathrm{cl}}}
\newcommand{\vinfty}{\ensuremath{\vel_\infty}}

\newcommand{\Mdot}{\ensuremath{\dot{M}}}
\newcommand{\zav}[1]{\left(#1\right)}

\begin{document}

\title{Validity of clumping approximations for mass-loss rates
determination}
\author{Ji\v{r}\'{\i} Kub\'at$^1$, Brankica \v{S}urlan$^{2,1}$}
\affil{$^1$Astronomick\'y \'ustav AV \v{C}R, Fri\v{c}ova 298, 251 65
Ond\v{r}ejov, Czech Republic}
\affil{$^2$Matematicko-fyzik\'aln\'{\i} fakulta UK, Ke Karlovu 3, 121 16
Praha 2}

\begin{abstract}
Clumping in stellar winds of hot stars is a possible consequence of
radiati\-ve-acoustic instability appearing in solutions of
radiative-hydrodynamical equations. However, clumping is usually
included to stellar atmosphere modeling and radiative transfer
calculations in a highly approximate way via a global free parameter
called the clumping factor. Using different values of clumping factors
many researchers succeeded to fit the observed spectra better and to
correct empirical mass loss rates. This usually leads to a conclusion
that the stellar wind is clumped.
To understand how clumping may influence theoretical predictions of
mass-loss rates, different clumping properties have to be taken into
account.
If clumping appears already below the critical point, the mass-loss
rates is changed.
\end{abstract}

\section{Introduction}

Many inhomogeneous structures in different astronomical objects have
been directly observed, in supernova remnants, planetary nebulae,
or just recently structures in the envelope of the red
supergiant $\alpha$~Ori \citep{2011A&A...531A.117K}.
Consequently, there is no reason to assume that the winds of hot stars
are perfectly homogeneous.

However, there are no such direct observations of wind structures
available.
The presence of clumping was inferred from line profile variability 
\citep[e.g.][]{1998ApJ...494..799E, 1999ApJ...514..909L,
2005A&A...440.1133M,2008cihw.conf.....H}.
Indirect evidence of clumping comes from X-ray observations
\citep{2004A&A...422..675O}.
Another evidence of clumping follows from detection of X-ray flaring in
Vela X-1, caused probably by wind clumps falling onto a neutron star
\citep{2010A&A...519A..37F}.

The generally accepted origin of clumping in hot star winds is the
radiative-acoust\-ic instability, supported by several
numerical simulations (see the review by Sundqvist, these proceedings).
An alternative explanation using adiabatic fluctuations
\citep{1997ApJ...482L.179C} did not attract attention of researchers.
However, recent interesting idea of the influence of subphotospheric
convection \citep{2009A&A...499..279C} offers a possibility of existence
of clumps in the part of the wind below the critical point, where the
numerical wind simulations do not predict it
\citep[e.g.][]{1995A&A...299..523F, 2002A&A...381.1015R}.
It seems that the most probable clumping scenario consists of multiple
clump formation mechanisms.
Although there is a lot of unknown about clump formation, distribution,
and properties, to find some kind of description of clumping is very
important.

\section{Description of clumping}

\subsection{Void interclump medium}

Due to missing direct observation of wind clumps the detailed treatment
of clumping is usually restricted to using free parameters.
This approach was previously used for the description of condensations
in nebulae by \cite{1959ApJ...129...26O} for the case of
the Orion nebula, where they assumed void intercondensation medium and
used the size of condensations as the free parameter, since the
observational data available to them did not allow to resolve details.
Contemporary advanced observational techniques made the usage of the
free parameter unnecessary.

The assumption of the void interclump medium is a common one for the
description of a clumped wind.
In other words it means that all matter is concentrated to clumps in vacuum.
Such medium can be described by means of one free parameter, which may
be either the {\em (volume) filling factor}
\begin{equation}\label{filfac}
f = \frac{V_\mathrm{clumps}}{V_\mathrm{wind}},
\end{equation}
a fractional volume which contains
material at higher density
\citep[referred to as the filling fraction by][]{2006ApJ...648..565O},
or the {\em clumping (correction) factor}
\begin{equation}\label{clfac}
D=C_c=f_\mathrm{cl} =
\frac{\langle\rho_\mathrm{clumps}\rangle}{\langle\rho_\mathrm{wind}\rangle}
= \frac{1}{f}.
\end{equation}
Just one of these two adjustable parameters is sufficient.
However, there is no reason to assume that the parameter defined by
(\ref{filfac}) or (\ref{clfac}) is depth independent.
Its depth dependence follows already from the pioneering hydrosimulations
of \cite{1988ApJ...335..914O}.
\cite{1999ApJ...519..354H} introduced an expression for the depth
dependence of the filling factor (\ref{filfac})
\begin{equation}\label{fvar}
f(r)=\finfty+(1-\finfty)\exp{\left(-\frac{\vel(r)}{\velcl}\right)},
\end{equation}
where {\finfty} corresponds to \eqref{filfac} and {\velcl} is the
location in the wind where clumping becomes important.
Many radiative transfer calculations have been done using this
expression.
Recently, \cite{2006A&A...454..625P} tried to determine the depth
dependence of the filling factor from observations.

\subsection{Dense interclump medium}

The assumption of the void interclump medium simplifies the description
of clumping.
The non-void interclump medium was assumed already by
\citet{1981ApJ...250..645A}, however, it was abandoned by later studies.
Recently the idea of a dense interclump medium was revoked by
\cite{2008ApJ...685L.149Z}.
The effect of dense interclump medium was studied in detail by
\cite{2010A&A...510A..11S, 2011A&A...528A..64S} and \v{S}urlan et al.
(these proceedings).
Inclusion of the non-void interclump medium requires additional free
parameter $d$, which relates the interclump density to the clump density
or to the density of the smooth wind, respectively.

\subsection{Clump properties}

In most radiative transfer calculations with clumping it is assumed that
clumps are optically thin, which means that clumps are smaller than the
mean free path of photons.
This assumption is sometimes referred to as the ``microclumping''.
However, it is more natural to assume that clumps may be optically thick
in some frequencies.
These clumps are larger than the mean free path of photons, which may
happen both in continua and lines.
This situation is being referered to as the ``macroclumping''
\citep{2007A&A...476.1331O} or ``porosity'' \citep{2006ApJ...648..565O}.

Since no 3D hydrodynamical simulations of the wind are available, there
is no hint what the shape of clumps may be.
In the parametric treatment using the clumping factor nothing is
explicitly assumed about the clump shape and sizes.
In more detailed calculations different shapes were assumed, like
spheres (\v{S}urlan et al., these proceedings), cubes
\citep{2011A&A...526A..32M}, shell fragments
\citep{2004A&A...422..675O}.
On the other hand, it is not clear how important is the detailed clump
shape, probably more important factor is the distribution of
clumps due to the stochastic nature of clumping.

The common assumption of most calculations is a smooth velocity field both
inside and outside clumps.
However, as follows from hydrodynamical simulations, assuming
nonmonotonic field inside clumps is probably closer to reality
\citep[see][]{2008cihw.conf..121O}.
This generalization was studied in more detail by
\citet[][and these proceedings]{2010A&A...510A..11S,2011A&A...528A..64S}
and \v{S}urlan et al. (these proceedings).
Note, however, that inhomogeneous velocity field can affect only
spectral lines.

\section{Influence of clumping on empirical mass-loss rates}

Mass-loss rates for particular stars are usually determined from
comparison of model emergent radiation with observations.
The emergent radiation is usually calculated assuming (i.e. for {\em
given}) velocity $\vel(r)$ and density $\rho(r)$ structure (and,
consequently, the mass-loss rate {\Mdot} and the terminal velocity
\vinfty).
The velocity structure is usually assumed to obey the so-called
$\beta$-velocity law
$\vel={\vinfty} \zav{1-{R_*}/{r}}^{\beta}$,
where $R_*$ is the stellar radius and $\beta$ is a free parameter.
Note that this type of dependence was already derived by
\cite{1926MNRAS..86..459M} and \cite{1934MNRAS..94..522C} with
$\beta=0.5$.
Today's estimates of this value are a bit higher.

The mass-loss rates are usually determined using different parts of the
stellar spectrum, like radio flux, infrared flux, the H$\alpha$ line, UV
resonance lines, and using synthetic spectra from model atmospheres.
Various determination methods were reviewed by
\cite{2008A&ARv..16..209P}.
Different diagnostics result in different mass-loss rates
\citep[e.g.][]{2003ApJ...595.1182B, 2006ApJ...637.1025F}.
Clumping in stellar winds is the suggested and promising way out from
this problem.

\cite{1991A&A...247..455H} tested clumping by artificial periodic
variation of $\rho(r)$.
Later \cite{1999ApJ...519..354H} introduced a depth variable filling
factor \eqref{fvar}, which became a common method for inclusion of
clumping in line formation calculations.
Since many studies found that mass-loss rates with clumping taken into
account in this way fit observations better, then the conclusion about
the presence of clumping was drawn.

\subsection{Unsaturated resonance lines}

Saturated resonance ultraviolet lines fit theoretical spectra usually
well, but their sensitivity to changes of wind parameters is relatively
low.
On the other hand, if lines are not saturated, they become a sensitive
diagnostics tool.
This happens for resonance lines of less abundant ions like
\ion{S}{iv}, \ion{S}{v}, \ion{Si}{iv}, or  \ion{P}{v}.
The case of the \ion{Si}{iv} resonance line (1393.76\,\AA, 1402.77\,\AA)
was recently studied by \cite{2010A&A...521L..55P} and from similar line
strengths of the resonance doublet they concluded that the wind is
clumped, since in the opposite case the line ratio should be~2.

The case of the resonance lines of \ion{P}{v} (1117.98\,\AA,
1128.01\,\AA) is more famous.
The possible importance of this doublet for mass-loss rates determination
was pointed out by \cite{2002ApJ...579..774C}.
\cite{2003ApJ...586..996M} and later  \cite{2006ApJ...637.1025F} found a
discrepancy in mass-loss rates determination and concluded that either
the mass-loss rates determined from \ion{P}{v} resonance lines are wrong
(which may be corrected by inclusion of wind clumping) or the
abundance of \ion{P}{v} is lower.

To test the abundance of \ion{P}{v}, \cite{2009MNRAS.394.2065K} studied
the NLTE ionization balance of phosphorus in the wind.
Using the code of \cite{2004A&A...417.1003K} they calculated the
phosphorus ionization balance both with and without X-rays and found
that the changes were insignificant.
Consequently, the abundance of \ion{P}{v} is not lowered by presence of
X-rays, which supports the clumping hypothesis.
Howerver, \cite{2010ApJ...711L..30W} suggested that XUV radiation may be
important for phosphorus ionization balance, detailed NLTE calculation
is still missing, however.

\section{Influence of clumping on predicted mass-loss rates}

While it is basically known how clumping (treated in the parametric way
using the volume filling factor) influences the empirical determination
of mass-loss rate for {\em given} velocity and density structure, for
the case of mass-loss rate predictions the situation is much less clear.

Solution of hydrodynamic equations for given basic stellar parameters
(e.g. the effective temperature $\Teff$, luminosity $L_*$, and radius
$R_*$) and radiation at the lower boundary gives the density $\rho(r)$
and velocity $\vel(r)$ structure of the wind.
The values of the mass loss rate $\Mdot$ and terminal velocity
$\vinfty$ follow directly from the determined structure, they are usually
called predicted.
The principal question is, how clumping influences predicted values of
$\Mdot$ and $\vinfty$.

This was studied by \citet{2008IAUS..252..283Ka,2008cihw.conf..111K}
using the stationary hydrodynamic code for NLTE stellar winds
\citep{2004A&A...417.1003K}, which consistently calculates the radiative
force without using the parameterization by the CAK parameters  $k$,
$\alpha$, and $\delta$).
Clumping was treated  in an approximate way ($\rho_\mathrm{clump}=C_c
\langle\rho\rangle$ -- cf. Eq.~\ref{clfac}), void interclump medium was
assumed.
This assumption modified the opacity and emissivity in a different way
for lines and continua, consequently the radiative force was modified.
Both optically thin and optically thick clumps were tested.

They found that if clumping starts below the critical point, then the
mass-loss rate increases.
If clumping starts above the critical point, then the mass-loss rate
does not change, but the terminal velocity increases.
Clumps larger than the  Sobolev length result in the decrease of
the mass-loss rate.
Clumping also influences the ionization balance, which has a strong
impact on the line force and, consequently, wind acceleration.
\cite{2011A&A...526A..32M} studied the effect of different portions of
clumped and unclumped parts of the wind and different clump sizes
(assuming that clumps are cubes).
They used the Monte Carlo wind code of
\cite{2000A&A...362..295V,2001A&A...369..574V} and found that optically
thin clumps increase the mass-loss rate, whereas optically thick clumps
decrease the mass loss rate.

\section{Conclusions}

The problem of \ion{P}{v} mass-loss rate determination has not been
satisfactorily solved yet.
While X-rays do not seem to alter the ionization balance of \ion{P}{v},
the case of XUV radiation has still to be tested by detailed consistent
NLTE calculations.

The dependence of mass-loss rates on clumping has still not been
analysed in a big detail yet.
Available results show that clumping below the critical point may have
strong influence on predicted mass-loss rates.
Although results of hydrodynamical simulations do not support clumping
below the sonic point, there is a possibility of creating clumps by
other mechanisms.
Detailed hydrodynamical calculations are needed.

\acknowledgements 
GA \v{C}R grants 205/08/H005 and 205/08/0003.

\bibliographystyle{asp2010}
\bibliography{kubat}

\end{document}